\documentstyle[aps,twocolumn,epsf]{revtex}

\def\be{\begin{equation}}
\def\ee{\end{equation}}

\newcounter{fig}

\catcode`\@=11   

\newcommand{\fcaption}[1]{\vspace{1ex}   
        \refstepcounter{figure}   
        \setbox\@tempboxa = \hbox{\footnotesize {\bf Fig.~\thefigure.} #1}   
        \ifdim \wd\@tempboxa > 8cm   
           {\begin{center}   
        \parbox{8cm}{\footnotesize\baselineskip=8pt {\bf Fig.~\thefigure.} #1}   
            \end{center}}   
        \else   
             {\begin{center}   
             {\footnotesize {\bf Fig.~\thefigure.} #1}   
              \end{center}}   
        \fi}

\begin{document}
\title{L\'{e}vy Flights from a Continuous-Time Process}
\author{I.M. Sokolov}
\address{Theoretische Polymerphysik, Universit\"{a}t Freiburg, Hermann-Herder-Str. 3,
\\D-79104 Freiburg im Breisgau, Germany}

\date{\today}
\maketitle

\begin{abstract}
The L\'{e}vy-flight dynamics can stem from simple random walks in a system
whose operational time (number of steps $n$) typically grows superlinearly
with physical time $t$. Thus, this processes is a kind of continuous-time
random walks (CTRW), dual to usual Scher-Montroll model, in which $n$ grows
sublinearly with $t$. The models in which L\'{e}vy-flights emerge due to a
temporal subordination let easily discuss the response of a random walker to
a weak outer force, which is shown to be nonlinear. On the other hand, the
relaxation of en ensemble of such walkers in a harmonic potential follows a
simple exponential pattern and leads to a normal Boltzmann distribution. The
mixed models, describing normal CTRW in superlinear operational time and
L\'{e}vy-flights under the operational time of subdiffusive CTRW lead to
paradoxical diffusive behavior, similar to the one found in transport on
polymer chains. The relaxation to the Boltzmann distribution in such models
is slow and asymptotically follows a power-law.

\bigskip 

\end{abstract}

\section{Introduction}

Random walk processes leading to subdiffusive or superdiffusive behavior are
adequate for describing various physical situations. Thus, the
continuous-time random walk (CTRW) model of Scher and Montroll \cite{ScheM}
was a milestone in understanding of photoconductivity in strongly disordered
and glassy semiconductors, while the L\'{e}vy-flight models \cite{KlaShleZu}
are adequate for description of transport in heterogeneous catalysis \cite
{Bych}, self-diffusion in micelle systems \cite{OBLU}, reactions and
transport in polymer systems under conformational motion \cite{Sok0},
transport processes in heterogeneous rocks \cite{KBZS}, and for description
of behavior of dynamical systems \cite{Sporad}. The closely related models
appear in description of economic time series \cite{Trunk}. The
L\'{e}vy-related statistics were observed in hydrodynamic transport \cite
{Swinney}, and in the motion of gold nanoclusters on graphite \cite{Clusters}%
. The mixed models were proposed , in which the slow temporal evolution
(described by Scher-Montroll CTRW) is combined with the possibility of
L\'{e}vy-jumps, so that in general both sub- or superdiffusive behavior can
arise \cite{Fogedby1}.

The continuous-time random walks (CTRW) first introduced by Montroll and
Weiss \cite{MoWe} correspond to a stochastic model in which steps of a
simple random walk take place at times $t_{i}$, following some random
process with non-negative increments: $\tau _{i}=t_{i}-t_{i-1}\geq 0$. In a
mathematical language one says that CTRW is a process subordinated to random
walks under the operational time defined by the process $\left\{
t_{i}\right\} $. It is typically thought that a CTRW-scheme alone can not
describe any superdiffusive process, so that the introduction of very long
jumps is an inevitable part of building a model leading to superdiffusive
behavior.

Let us first discuss a typical CTRW approach. Let us consider a
one-dimensional situation under which a particle from time to time makes a
jump to a neighboring lattice site separated from the initial one by a
distance $a$. The time $\tau $ between the two jumps is distributed
according to some waiting-time distribution, represented by the probability
density function (PDF) $p(\tau )$. If the mean waiting time $\bar{\tau}$
exists, the particle's behavior is diffusive, with diffusion coefficient $%
D=a^{2}/2\bar{\tau}$. If the corresponding moment diverges, the particle's
behavior becomes subdiffusive, with $\left\langle r^{2}(t)\right\rangle
\propto t^{\alpha }$, with $\alpha <1$ depending on the PDF $p(\tau )$. The
subdiffusive behavior is indicated by vanishing of the diffusion coefficient 
$D$. It seems impossible to obtain within this scheme any type of a
superdiffusive behavior unless one allows for infinitely long jumps with $%
\left\langle a^{2}\right\rangle \rightarrow \infty $. The superdiffusive
behavior is indicated by divergence of the diffusion coefficient $D$. If $%
\left\langle a^{2}\right\rangle $ stays finite this can be the case only if $%
\bar{\tau}$ vanishes. Since $\tau >0$ and $\bar{\tau}=\int_{0}^{\infty }\tau
p(\tau )d\tau ,$ vanishing of the mean waiting time means that $p(\tau
)=\delta (\tau )$, a  marginal, degenerate situation.

On the other hand the consideration presented above shows only that the
waiting-time distribution is not an adequate tool for description of
superdiffusive CTRW. In what follows we show that superdiffusive CTRW with
bounded step lengths are just as possible as the subdiffusive ones. Our
considerations will be rather formal and do not follow from any particular
physical model. On the other hand, the fact that L\'{e}vy-flights can stem
from a process subordinated to simple random walks has many important
implications. Thus, as we proceed to show, the fast dynamics of a free
process can coexist in such models with
simple exponential relaxation to a normal Boltzmann equilibrium
distribution, if the behavior of an ensemble of random walkers under
restoring force is considered. This shows that the relation between L\'{e}vy
dynamics and the nonextensive thermodynamics described by nonclassical
entropy functions is much looser than typically assumed.

The combinations of the superdiffusive L\'{e}vy-flights with the typical
CTRW operational time leads to paradoxical diffusion behavior, having some
parallels in transport on polymer chains. Moreover, the existence of a
subordination model leading to L\'{e}vy flights can be useful in
understanding of statistical implications of the processes described by
fractional generalizations of diffusion and Fokker-Planck equations \cite
{Hilfer,MKS,BarMetzKla}.

The article is organized as follows: In Sec. 2 we discuss general properties
of subordinated random processes. In Sec.3 and 4 the processes subordinated
to symmetric and asymmetric random walks are considered, these leading to
symmetric and asymmetric L\'{e}vy-flights. The dualism between the
L\'{e}vy-flights and the Scher-Montroll CTRW is discussed in Sec.5. Sections
6 and 7 discuse the models leading to paradoxical diffusion behavior. The
relaxation to equilibrium is considered in Sec. 8.

\section{The subordination of random processes}

As already mentioned, a Scher-Montroll CTRW process is a simple random walk
whose steps take place at times $t_{i}$ governed by a random process with
nonnegative independent increments, so that

\begin{equation}
P(x,t)=\sum_{n}P_{RW}(x,n)p_{n}(t),  \label{CTRW}
\end{equation}
where $P_{RW}(x,n)$ is a probability distribution to find a random walker at
point $x$ after $n$ steps (i.e. the binomial distribution), and $p_{n}(t)$
is the probability to make exactly $n$ steps up to time $t$. For both $t$
and $n$ large, when the binomial distribution can be approximated by a
Gaussian one, and when the corresponding sum can be changed to an integral,
Eq.(\ref{CTRW}) reads: 
\begin{equation}
P(x,t)\simeq \int_{0}^{\infty }\frac{1}{\sqrt{2\pi n}}\exp \left( -\frac{%
x^{2}}{2n}\right) p_{t}(n,t)dn.  \label{Cont}
\end{equation}
In a classical Scher-Montroll CTRW $p_{t}(n,t)$ corresponds to a random
process in which $n$ typically grows sublinearly in $t$. Thus, the overall
process is subdiffusive.

Note that a description of CTRW-process given by Eq.(\ref{Cont}) is an
example of \textit{subordination}, see Sec. X.7 of Ref.\cite{Feller}: If $%
\left\{ X(T)\right\} $ is a Markov process with continuous transition
probabilities and $\left\{ T(t)\right\} $ a process with non-negative
independent increments, then $\left\{ X(T(t))\right\} $ is said to
subordinate to $\left\{ X(t)\right\} $ using the operational time $T$. In
this case 
\begin{equation}
P(x,t)=\int_{0}^{\infty }P_{x}(x,T)p_{T}(T,t)dT.  \label{Sub1}
\end{equation}
In what follows we call the integral transform, Eq.(\ref{Sub1}) a
subordination transformation, changing from time scale $t$ to a time-scale $T
$. For example, in the Scher-Montroll case the operational time of a system
is given by the number of steps of the RW, and is a random function of the
physical time $t$ whose typical value grows sublinearly in $t$.

The operational time can also grow superlinearly with $t$. Such a process
can not be described by a waiting-time distribution, and needs a
complimentary description. Let us consider a random process, where the 
\textit{density} of events fluctuates strongly. Let us subdivide the time
axis into intervals of duration $\Delta t$ and let us consider the number $n$
of jumping events within each interval. The value $\rho =n/\Delta t$ defines
the density of jump events. Now, if the mean density of events exists, its
inverse gives us exactly the mean waiting time of a jump, and a process
described by a finite density of events is a normal diffusive one. The
divergence of a mean waiting time (like in Scher-Montroll CTRW) correspond
to vanishing density. On the other hand, if one considers a strongly
fluctuating density $\rho (t)$ whose first moment diverges, the mean waiting
time vanishes and a process that subordinates a random walk process under
such operational time can be superdiffusive. At longer times, the
distribution of the number of events tends to one of the L\'{e}vy-stable
laws: the typical number of events can grow superlinearly in time. A simple
example of such process was already known to Feller, see Chap. X.7 of Ref.%
\cite{Feller}. He considers a process subordinated to simple random walks
under the operational time governed by a fully asymmetric L\'{e}vy stable
law of index 1/2. The corresponding PDF at time $t$ is given by 

\begin{eqnarray}
P(x,t) &=&\int_{0}^{\infty }\frac{1}{\sqrt{2\pi n}}\exp \left( -\frac{x^{2}}{%
2n}\right) \frac{t}{\sqrt{\pi }n^{3/2}}\exp \left( -\frac{t^{2}}{2n}\right)
dn  \nonumber \\
&=&\frac{t}{\pi (t^{2}+x^{2})},  \label{Fellerex}
\end{eqnarray}
i.e. is a Cauchy L\'{e}vy-flight.

Let us now discuss a simple analogy describing the relation between the
Scher-Montroll CTRW and L\'{e}vy-flights. This analogy makes clear many of
the findings we are going to discuss below. Imagine a physical clock
producing ticks following with frequency 1, which govern the behavior of a
random walker. Imagine a switch situated at $0$, so that returning to the
origin, the walker can trigger some physical process (the analogy with the
Glarum model of relaxation, Ref.\cite{Glarum}, is evident!). The times
between the subsequent returns are distributed according to a fully
asymmetric L\'{e}vy stable law of index 1/2 used in a previous example.
Imagine now another random walker performing its motion (a step per physical
unit time) independently from the first one. Imagine a movie camera, taking
frame-per-frame pictures of the positions of this second random walker at
the moments when the first walker is at the origin and thus triggers the
switch. Watching the movie taken by the camera, we immediately recognize
that the second walker performs the Cauchy L\'{e}vy-flights. Imagine, that a
clock is posed in a frame and also filmed. In this case its image will show
exactly the operational time of the system; the spectator's watch measures
the physical time. Imagine an opposite situation: the first walker triggers
the motion of the second one, and the camera is triggered by the physical
clock, as a normal movie camera is. The process we recognize at the film is
then the Scher-Montroll CTRW. We can take a Scher-Montroll movie also using
another trick (which cannot be performed in a real time, but needs a record
of return times). Let us take a record of subsequent return times of a first
random walker (numbers $n_{1},$ $n_{2},$ ...) and trigger our camera in such
a way that it makes $n_{1}$ frames during the first second, $n_{2}$ frames
during the second one, etc. If we film a normal random walker with a camera
prepared in such a way, the movie will show us the Montroll-Weiss CTRW. An
image of the physical clock will again show the operational time of the
system, and again, looking at his watch, the spectator can measure the
physical time between two events.

Let us use our camera triggered by returns of a random walker to film other
processes taking place in the outer world. The film which is watched
afterwards under constant speed shows us a \textit{possible} world: The
causality relations and thermodynamical time arrow are those of our usual
world. On the other hand, a movie of a world undergoing continuous
evolution, in which ''\textit{natura non facit saltus''} holds, will show us
a revolutionary world of ''great leaps'' and abrupt changes (but following
the same logics of development). The second camera (fed by a prescribed $n$%
-sequence) will show us the world of almost full stagnation seldomly
interrupted by a bounded, local movement, a world developing in a slow time
of old Asiatic despoty. We shall keep this analogy in mind when discussing
the physical implications of subordination.

Let us consider a system which evolves according to a Markovian dynamics and
whose state tends to a normal Boltzmann equilibrium under relaxation. In a
system under action of outer forces, the transition probabilities between
the states of the system (sites $i$ between which the random walk takes
place) which are characterized by their energies $E_{i}$, are not
independent. They are connected through the corresponding Boltzmann-factors,
so that in equilibrium during any period of time $\Delta t$ the mean numbers
of forwards and of backwards jumps between any two sites $i$ and $j$ fulfill
the condition 
\begin{equation}
n_{ij}(\Delta t)/n_{ji}(\Delta t)=\exp \left[ \left( E_{i}-E_{j}\right)
/kT\right] ,  \label{DeB}
\end{equation}
where $k$ is the Boltzmann constant and $T$ is the system's temperature. The
condition Eq.(\ref{DeB}) guarantees detailed balance in equilibrium,
independentl of what the real dynamics of a system is. For simple RWs, where
only transitions between the neighboring states are allowed, the
corresponding transition rates with respect to the operational time of the
system can be introduced. For a random a walker moving under the influence
of a weak constant force $F$ the probabilities of the forward and backward
jumps per unit time $w_{+}$ and $w_{-}$ are connected through $%
w_{+}/w_{-}=\exp (Fa/kT)$. The Markovian nature of RW then leads to the fact
that the values of $w_{+}$ and $w_{-}$ do not depend on whether the system
is in equilibrium or not. For $F$ small one can take, say, $%
w_{+}=w_{0}(1+Fa/kT)$ and $w_{-}=w_{0}(1-Fa/kT)$ with $w_{0}=1/2\tau $.

Note that subordination, describing a transition from a physical time to an
operational time of the system, does not change its equilibrium properties.
Such subordination can be considered as random modulation of the transition
rate $w_{0}$ by some independent process (say closing and opening the
channels), and is fully irrelevant for thermodynamics (i.e. thermostatics)
of the system. On the other hand, it strongly influences its kinetics, so
that a question can be posed, what kinds of kinetics are compatible with the
relaxation to a normal Boltzmann distribution under arbitrary subordination
transformation of time. We address this question in Sec.8, after the free
diffusion properties of superdiffusive CTRW will be discussed.

\section{Symmetric L\'{e}vy flights from CTRW}

Let us first concentrate on the symmetric random walk case. Let us consider
a random process in which the number of events per given time is unbounded
and follows, for example, a power-law distribution, $p_{n}(t)\propto
tn^{-1-\alpha }$ with $0<\alpha \leq 1$ (this corresponds to the typical
number of events scaling as $n\propto t^{1/\alpha }$). Let us find the
asymptotic behavior of $P(x,t)$ for $t$ large. Since the jumps during
different intervals are uncorrelated, the PDF of $n$ for longer times
converges to a fully asymmetric L\'{e}vy-stable law 
\begin{equation}
p(n,t)\simeq t^{-1/\alpha }L(n/t^{1/\alpha };\alpha ,\gamma )
\end{equation}
with the asymmetry parameter $\gamma =-\alpha $ (here the values of $\gamma
=\pm \alpha $ correspond to the strongly asymmetric PDF that vanish
identically for large positive (negative) $x$ values , while $\gamma =0$
corresponds to symmetric distributions; the notation in one of Ref.\cite
{Feller}). Note that the Fourier-transforms of L\'{e}vy-stable laws are
known: up to the translation $P(k,t)$ is equal to 
\begin{equation}
f(\kappa )=\exp \left[ -\left| \kappa \right| ^{a}e^{i\pi \gamma /2}\right]
\end{equation}
(for $0<\alpha <2$, $\alpha \neq 1$). The PDF is a real function, thus $%
f(\kappa )=f^{*}(-k)$. The corresponding function is analytical everywhere
except for $\kappa =0$, so that the PDF is given by 
\begin{equation}
L(x;\alpha ,\gamma )=\frac{1}{\pi }\mbox{Re}\int_{0}^{\infty }e^{-ix\zeta
-\zeta ^{\alpha }e^{i\pi \gamma /2}}d\zeta.  \label{Levy0}
\end{equation}
From Eq.\ref{Levy0} the series expansions for $L(y;\alpha ,\gamma )$ follow,
see Sec. XVII.6 of Ref.\cite{Feller}. In the case $\alpha <1$ one can move
the path of integration to the negative imaginary axis (since the integrand
tends to zero when $\mbox{Im}\zeta \rightarrow -\infty $ due to the
dominance of the linear term), which allows then for elementary integration
after Taylor-expansion of $\exp (A\zeta ^{\alpha })$. For $1<\alpha <2$ this
dominance is no more the case, but the integrand still vanishes for $\mbox{Im%
}\zeta \rightarrow -\infty $ in the case of symmetric distributions, while $%
(-i\left| \zeta \right| )^{\alpha }=\left| \zeta \right| ^{a}(\cos \frac{\pi 
}{2}\alpha -i\sin \frac{\pi }{2}\alpha )\rightarrow -\infty $ for $1<\alpha
<2$. Thus the series which represents L\'{e}vy distributions for $0<\alpha
<1 $ and symmetric L\'{e}vy distribution also for $1<\alpha <2$ reads:

\begin{equation}
L(y;\alpha ,\gamma )=\frac{1}{\pi y}\sum_{k=1}^{\infty }(-1)^{k}\frac{\Gamma
(k\alpha +1)}{k!}\sin \left( \frac{k\pi }{2}(\gamma -\alpha )\right)
y^{-\alpha k}  \label{Lest}
\end{equation}
In general the L\'{e}vy-stable laws for $1<\alpha <2$ are given by another
expansion, 
\begin{equation}
L(y;\alpha ,\gamma )=\frac{1}{\pi y}\sum_{k=1}^{\infty }(-1)^{k}\frac{\Gamma
(1+k/\alpha )}{k!}\sin \left( \frac{k\pi }{2}(\gamma -\alpha )\right) y^{-k}
\end{equation}
which also holds for asymmetric laws.

One can easily obtain the form of $x$-distributions by immediate
integration: using Eq.(\ref{Cont}) and a scaling form of a
L\'{e}vy-distribution 
\begin{equation}
p(x,t)\simeq \int_{0}^{\infty }\frac{1}{\sqrt{2\pi n}}\exp \left( -\frac{%
x^{2}}{2n}\right) L(\frac{n}{t^{1/\alpha }};\alpha ,-\alpha )\frac{dn}{%
t^{1/\alpha }}.
\end{equation}
Using Eq.(\ref{Lest}) and performing a term-by-term integration, we arrive
to the series of integrals of the form 

\begin{eqnarray}
I_{\mu }(\zeta ) &=&\int_{0}^{\infty }\frac{1}{\sqrt{2\pi \xi }}e^{-\frac{%
\zeta ^{2}}{2\xi }}\xi ^{-\mu }d\xi   \nonumber \\
&=&\frac{1}{\sqrt{2\pi }}\left( \frac{\zeta ^{2}}{2}\right) ^{1/2-\mu
}\Gamma (\mu -1/2).  \label{v=0}
\end{eqnarray}
For integral of the $k$-th term in Eq.(\ref{Lest}) we have $\mu =1+\alpha k$%
. Let us concentrate first on the case $0<\alpha <1$. Using well-known
relations for $\Gamma $-function: $\Gamma (z+1)=z\Gamma (z)$ (Eq. (6.1.15)
of Ref.\cite{Abrasteg}) and $\Gamma (2z)=(2\pi )^{-1/2}2^{2z-1/2}\Gamma
(z)\Gamma (z+1/2)$ (Eq. (6.1.18) of \cite{Abrasteg}) we get 
\begin{equation}
p(\zeta )=\frac{1}{\pi }\sum_{k=1}^{\infty }(-1)^{k}\frac{\Gamma (2k\alpha
+1)}{k!}\sin \left( -k\pi \alpha \right) \left( \frac{\sqrt{2}}{\zeta }%
\right) ^{-2\alpha -1},
\end{equation}
which represents a series expansion for a symmetric L\'{e}vy-stable law of
index $2\alpha $, Eq.(\ref{Lest}), for the scaled variable $\zeta /\sqrt{2}$%
. This corresponds to a form $p(x,t)=t^{-1/2\alpha }L(x/\sqrt{2}t^{2\alpha
};2\alpha ,0)$ of the $x$-distribution.

We note that taking Fourier-transform of the both parts of for symmetric
RWs, 
\begin{equation}
L(ax,2\alpha ,0)=\int_{0}^{\infty }\frac{1}{\sqrt{2\pi n}}\exp \left( -\frac{%
x^{2}}{2n}\right) L(n;\alpha ,-\alpha )dn,
\end{equation}
where $a$ is an unimportant scaling factor, we get: 
\begin{equation}
\exp (-A\left| k\right| ^{2\alpha }t)=\int_{0}^{\infty }\exp \left(
-k^{2}n\right) L(n;\alpha ,-\alpha )dn,  \label{Lapl}
\end{equation}
which holds for any real $k$ (i.e. for any positive $k^{2}$), where $A\,$is
a number factor. This gives us a general expression for a Laplace-transform
of an asymmetric L\'{e}vy-distribution with $\alpha <1$: A Laplace-transform
of $L(n,\alpha ,-\alpha )$ is $\exp (-A\left| k\right| ^{\alpha }t)$. From
this fact an important result follows: 
\begin{equation}
L(ax;\alpha \beta ,0)=\int_{0}^{\infty }n^{-1/\beta }L(x/n^{1/\beta };\beta
,0)L(n;\alpha ,-\alpha )dn:  \label{Subord}
\end{equation}
A L\'{e}vy distribution with index $\alpha \beta $ is subordinated to a
L\'{e}vy-distribution with index $\beta <\alpha $ under the operational time
given by an asymmetric L\'{e}vy law of index $\alpha <1$. To see this,
consider the characteristic functions of both sides of Eq.(\ref{Subord}) and
use Eq.(\ref{Lapl}).

\begin{equation}
\exp (-A\left| k\right| ^{\alpha \beta })=\int_{0}^{\infty }e^{-\left|
\kappa \right| ^{\beta }n}L(n;\alpha ,-\alpha )dn,  \label{SBR}
\end{equation}
see Sec.X.7 of Ref.\cite{Feller}. Eq.(\ref{Lapl}) corresponds to a special
case of $\beta =2$ of Eq.(\ref{SBR}). The distributions $L(n;\alpha ,-\alpha
)$ thus coincide with inverse Laplace transforms of stretched-exponentials.
For example for $L(n;1/2,-1/2)$ one readily gets: 
\begin{equation}
p(n,t)={{\mathcal{L}}^{-1}} \left\{ \exp (-tu^{1/2})\right\} =\frac{t}{2\sqrt{\pi 
}n^{3/2}}\exp \left( -\frac{t^{2}}{4n}\right) \text{,}  \label{Smir}
\end{equation}
which differs only by a scale for the time-unit from a distribution used in
the example Eq.(\ref{Fellerex}).

\section{Asymmetric L\'{e}vy-flights}

Imagine a random walker moving under the influence of a weak constant force $%
F$. Such force introduces an asymmetry into the walker's motion, since the
probabilities of the forward and backward jumps, $w_{+}$ and $w_{-}$ are now
weighed with the corresponding Boltzmann-factors, $w_{+}/w_{-}=\exp (Fa/kT)$%
. For $F$ small one can take $w_{+}=1/2+Fa/2\tau kT$ and $w_{+}=1/2+Fa/2\tau
kT$. For $t$ large such random walks lead to the Gaussian distribution of
the particles' positions 
\begin{equation}
P_{RW}(x,t)=\frac{1}{\sqrt{2\pi n}}\exp \left( -\frac{(x-vn)^{2}}{2n}\right)
\end{equation}
whose center moves with a constant velocity $v=\mu F=Fa^{2}/2\tau kT$. Note
that our RW fulfil the Einstein's relation between the mobility $\mu $ and
diffusion coefficient $D$: $\mu =D/kT$. The PDF of a random process which
subordinates biased RW under an operational time following the asymmetric
L\'{e}vy-law is given by:

\begin{eqnarray}
P(x,t)\simeq \int_{0}^{\infty }\frac{1}{\sqrt{2\pi n}}\exp \left( -\frac{%
(x-vn)^{2}}{2n}\right)   \nonumber \\
\times L(\frac{n}{t^{1/\alpha }};\alpha ,-\alpha )\frac{dn}{%
t^{1/\alpha }}.  \label{Asymm}
\end{eqnarray}
Using the series expansion, Eq.(\ref{Lest}) and performing the term-by-term
integration leads to the series of the integrals of the type:

\begin{eqnarray}
I_{\mu }(\zeta ,\omega ) &=&\int_{0}^{\infty }\frac{1}{\sqrt{2\pi \xi }}e^{-%
\frac{(\zeta -\omega \xi )^{2}}{2\xi }}\xi ^{-\mu }d\xi   \nonumber \\
&=&\frac{2\exp (\zeta \omega )}{\sqrt{2\pi }}\left( \frac{\zeta ^{2}}{\omega
^{2}}\right) ^{1/4-\mu /2}K_{1/2-\mu }\left( \zeta \omega \right) 
\end{eqnarray}
for $\omega \neq 0.$ For integral of the $k$-th term in Eq.(\ref{Lest}) we
again have $\mu =1+\alpha k$. Let us concentrate first on the case $0<\alpha
<1$. For $\zeta \omega $ small, $v$ cancels (see the expansion 9.6.9 of Ref.%
\cite{Abrasteg}, $K_{\nu }(z)\simeq \frac{1}{2}\Gamma (\nu )(\frac{1}{2}%
z)^{-\nu }$ ($\nu >0$), note that $K_{-\nu }(z)=K_{\nu }(z)$), so that the
corresponding distribution tends to be a function of $\zeta $ only, it
coincides with one for $\omega =0$, Eq.(\ref{v=0}) so that a symmetric
L\'{e}vy-stable law of index $2\alpha $, Eq.(\ref{Lest}) emerges. On the
other hand, for $v\neq 0$ and $x$ large the overall distributions follow
from the expansion of $K$ for large values of the argument which reads: $%
K_{v}(z)\simeq \sqrt{\frac{\pi }{2z}}e^{-z}$ (Eq. 9.7.2 of Ref.\cite
{Abrasteg}). The corresponding integral then tends to $\frac{1}{\nu }\left(
\zeta /\omega \right) ^{-\mu }$, so that the corresponding PDF reproduces
the PDF of the density of events (up to rescaling). This last form is also
the asymptotic from corresponding to the behavior of Eq.(\ref{Asymm}) for
large $t$.

Hence, the distribution $P(x,t)$ tends to a fully asymmetric one of index $%
\alpha $ for $x$ and $t$ large. In this case the distribution shows scaling
with a scaling parameter $\xi =x/(vt)^{\alpha }$. We see that in this case
the motion under the influence of a constant force is superdiffusive, so
that $x\simeq (Ft)^{1/\alpha }$, and its dependence on the outer force is
nonlinear. Thus, the model shows a behavior that differs considerably from a
linear-response assumption of Refs.\cite{Fogedby1,Sune,FogQ}. This absence
of a linear response regime is parallel to the CTRW-findings \cite{ScheM}
(see Ref.\cite{BuGe} for a review) 
and mirrors the fact that only for normal diffusion a
sweep with constant velocity and a drift under a constant force result in
the same pattern of motion, see Ref. \cite{MKS}.

The case $\alpha =1/2$ again results in a closed expression: 

\begin{eqnarray}
P(x,t) &=&\int_{0}^{\infty }\frac{1}{\sqrt{2\pi n}}\exp \left( -\frac{%
(x-vn)^{2}}{2n}\right)  \\
&&\times \frac{t}{\sqrt{2\pi n^{3}}}\exp \left( -\frac{t^{2}}{2n}\right) dn 
\nonumber \\
&=&\frac{1}{\pi }\frac{vt}{\sqrt{x^{2}+t^{2}}}e^{vx}K_{1}\left( \sqrt{%
v^{2}\left( x^{2}+t^{2}\right) }\right)   \nonumber
\end{eqnarray}
(2.3.16.1 of Ref.\cite{BP}). For $v,x$ and $t$ small, the corresponding
distribution tends to a Cauchy-law. On the other hand, for $t$ large we can
take approximately: 
\begin{equation}
P(x,t)\approx \frac{1}{\sqrt{2\pi }}\frac{\sqrt{v}t}{\left(
x^{2}+t^{2}\right) ^{3/4}}e^{v\left( x-\sqrt{x^{2}+t^{2}}\right) }
\label{Asmpt}
\end{equation}
The second moment of this distribution diverges, but the position of the
maximum of $P(x,t)$, determining the typical particle position at time $t$,
tends to grow as $x_{\max }=\frac{2}{3}t^{2}$ for $t$ large. Thus, the
typical behavior of $x(t)$ under constant force is superlinear.

Note that in the case $1<\alpha <2$ the distribution of the particle's
displacement for the case $v=0$ will tend to a Gaussian, but in the case $%
v>0 $ it still tends to a fully asymmetric L\'{e}vy one. On the other hand,
in this case the distribution of the particle's position possesses the first
moment which grows linearly with time, thus the situation under $\alpha >1$
shows the linear response behavior. Since the second moment of the
distribution is absent, the fluctuations are strong, and the width of such
distribution is of the order of the typical value of $x$ itself.

\section{The dualism between the sub- and the superdiffusive CTRW}

There exists a clear dualism between the normal, subdiffusive CTRW and a
superdiffusive one. The corresponding concepts are illustrated in discrete
time by Fig.1, where we return to a situation discussed in Sec.2. Imagine a
clock producing ticks following with frequency 1, marking the physical time
of a system. Imagine a system which is triggered not by each tick of a
physical clock, but follows some waiting-time distribution, $\psi (n)$. This
means that after our random walker has jumped, the next jump will take place
after $n$ ticks of a clock, where the number $n$ is chosen according to a
power-law distribution, say $\psi (n)\propto (n+1)^{-1+\gamma }$. The number 
$n$ fluctuates strongly, so that the sequence of jumps (corresponding to a
randomly decimated sequence of ticks) shows lacunae of different duration.
Fig. 1a) shows a realization of such a sequence for the case $\gamma =0.75$.
The lacuna starting in the middle of Fig. 1a) at $t=54$ ends at $t=161$. The
mean number of jumps during the time $t$ grows sublinearly with $t$, namely
as $t^{3/4}$. Let us denote the corresponding subordination transformation
as time-expanding transformation (TET) of index $\gamma $. According to the
procedure described above, the corresponding sequence does not have any
intervals where the density of events is larger than one. The process
subordinated to random walks under such operational time (normal CTRW) is
subdiffusive.

Let us now consider the sequence of jumps of a walker as ticks marking
relevant time epochs of a system (i.e. associate each jump with a tick of a
physical clock). From this point of view, the ticks of initial clocks follow
extremely inhomogeneously, so that the number of such ticks within a
physical time unit varies according to $p(n)\propto (n+1)^{-1+\gamma }$.
Fig.1b) illustrates this situation: Here we took 100 jumps from the
realization shown in Fig.1a) and rescaled each of the corresponding time
intervals to the unit length. The ticks of initial clock (shown as bars)
follow inhomogeneously and show the intervals of high concentration (but no
lacunae). The number of such events grows superlinearly in time. The
corresponding subordination transformation will be called a ''time-squeezing
transformation'' (TST) of index $\gamma $. The process subordinated to
random walks under such operational time is superdiffusive and corresponds
to L\'{e}vy-flights. Note that both TST and TET are the probability
distributions $P(n,t)$ of the operational time $n$ for a given physical time 
$t$, i.e. are positive, integrable functions of $n$.

\begin{figure} \begin{center}   
  \fbox{\epsfysize=4cm\epsfbox{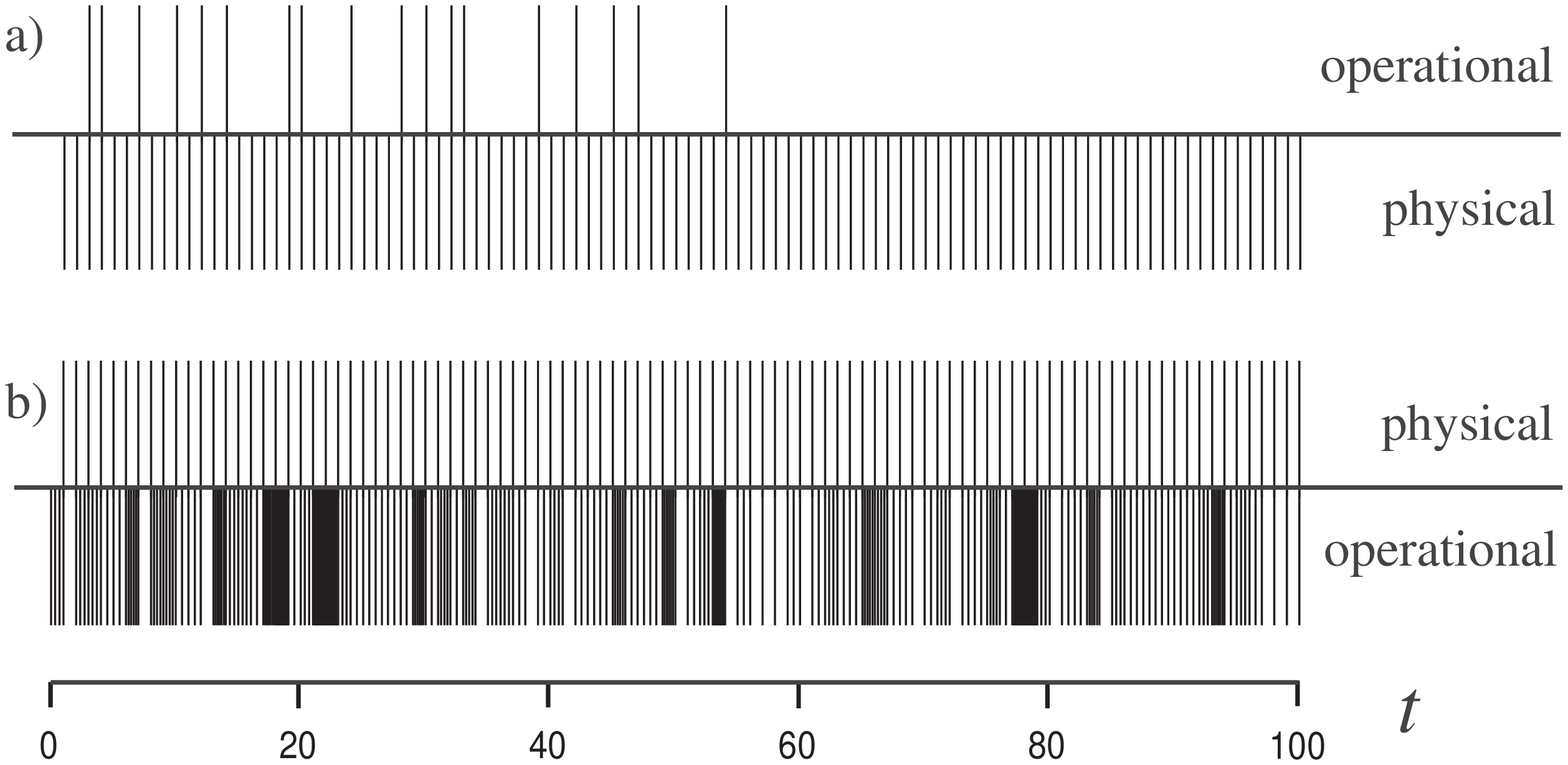}}   
   \fcaption{This figure illustrates the notion of the
operational time. a) The operational-time leading to the Scher-Montroll
CTRW. The ''ticks'' triggering the motion of the random walker (shown as
bars above the horizontal line) are taken from the set of the ''ticks'' of a
physical clock according to a waiting-time distribution $\psi (n)\propto
(n+1)^{-1+\gamma }$ (shown as bars below the line). The process shows long
lacunae, see text for details. b) We now declare each interval between the
two moves of the random walker for a new time unit. The initial ticks of the
physical clock follow now extremely inhomogeneously and show the intervals
of high condensation. This kind of operational time leads to
L\'{e}vy-flights.}   
  \label{tra}   
\end{center}\end{figure}

Let us now combine the two processes. For example, let us first generate the
superlinear sequence using algorithm described above (with $p(n)\propto
(n+1)^{-1+\gamma }$) and then decimate it randomly according to the
waiting-time distribution $\psi (n)\propto (n+1)^{-1+\gamma }$. The typical
number of events during time interval $t$ grows in this case linearly with $%
t $, but the corresponding sequence of events is extremely inhomogeneous,
showing both lacunae and accumulation intervals on all scales. This process
is shown in a bar-code-like picture in Fig.2. It will be discussed in more
detail in Sections 6 and 7. We can also proceed other way around, and apply
the transformations in the opposite way, namely, first generating a
sublinearly growing, lacunary operational time and then filling the lacunae
according to a L\'{e}vy distribution. As we proceed to show, these two ways
of constructing the event-time sets are not equivalent. The process
subordinated to RW under such inhomogeneous operational time is a kind of a
continuous-time L\'{e}vy-flight, and not a normal RW.

The example discussed above shows that transformations leading to sub- or
superlinear operational time behavior (dual to each other in the sense
described above) are not inverse of each other. Let us discuss a possibility
of a subordination transformation transforming a L\'{e}vy-stable
distribution of index $\beta $ (for example, a Gaussian distribution) into
one with ones with index $\gamma $, in the sense that 
\begin{equation}
L(ax;\gamma ,0)=\int_{0}^{\infty }n^{-1/\beta }L(x/n^{1/\beta };\beta
,0)S(n,t)dn,  \label{slow}
\end{equation}
where $S(n,t)$ is supposed to be a probability distribution of the number of
steps $n$ done up to time $t$. Taking Fourier-transform of both parts of Eq.(%
\ref{slow}) and changing to a variable $u=\left| k\right| ^{\beta }$ we get: 
\begin{equation}
\exp (-A\left| u\right| ^{\alpha })=\int_{0}^{\infty }e^{-un}S(n,t)dn
\label{ilap}
\end{equation}
with $\alpha =\gamma /\beta $. From Eq.(\ref{ilap}) it follows that $S(n,t)$
are the inverse Laplace transforms of stretched-exponentials $\exp
(-Au^{\alpha })$. Note that according to the Bernstein's theorem, a function 
$f(x)$ is a Laplace-transform of a probability distribution if and only if
it is completely monotone (i.e. it is infinitely differentiable and $%
(-1)^{n}f^{(n)}(x)\geq 0$ for all derivatives $f^{(n)}$) and $f(0)=1.$ The
last condition is always fulfilled. Note that according to Criterion 2
discussed on p.441 of vol.IIII of Ref.\cite{Feller} a function $%
f(x)=e^{-\psi (x)}$ is a completely monotone function if and only if $\psi $
is a positive function with a completely monotone derivative. In our case $%
\psi (x)=Au^{\alpha }.$ For $0<\alpha <1$ one has: $g(x)=\psi ^{\prime
}(x)=A\alpha u^{\alpha -1}>0$, and the higher derivatives (defined on the
interval $0<x<\infty $) are: $g^{\prime }(x)=A\alpha (\alpha -1)u^{\alpha
-2}<0,$ $g^{\prime \prime }(x)=A\alpha (\alpha -1)(\alpha -2)u^{\alpha -3}>0$%
, $g^{\prime \prime \prime }(x)=A\alpha (\alpha -1)(\alpha -2)(\alpha
-3)u^{\alpha -3}<0$, etc., so that $(-1)^{n}g^{(n)}(x)\geq 0$, and thus the
function $g$ is completely monotone. Thus $S(n,t)$ is a probability
distribution (namely the one which we have found above by explicit
calculation). On the other hand, for $\alpha >1$ the function $g(x)$ is not
completely monotone, so that $S(n,t)$ is \textit{not} a probability
distribution. Thus, there is no random process which defines the operational
time in such a way that the L\'{e}vy-flight of index $\alpha _{1}$ will be
transformed into a L\'{e}vy-flight with index $\alpha _{2}>\alpha _{1}$. The
absence of an inverse of a TST belonging to a class of subordination
transformations has a deep physical interpretation: a TST is a
coarse-graining procedure (see Fig.1): the information about the internal
steps of the process gets lost. One can not anticipate that the
transformation inverse to a coarse-graining belongs to the same class as the
direct one.

\begin{figure} \begin{center}   
  \fbox{\epsfysize=2.5cm\epsfbox{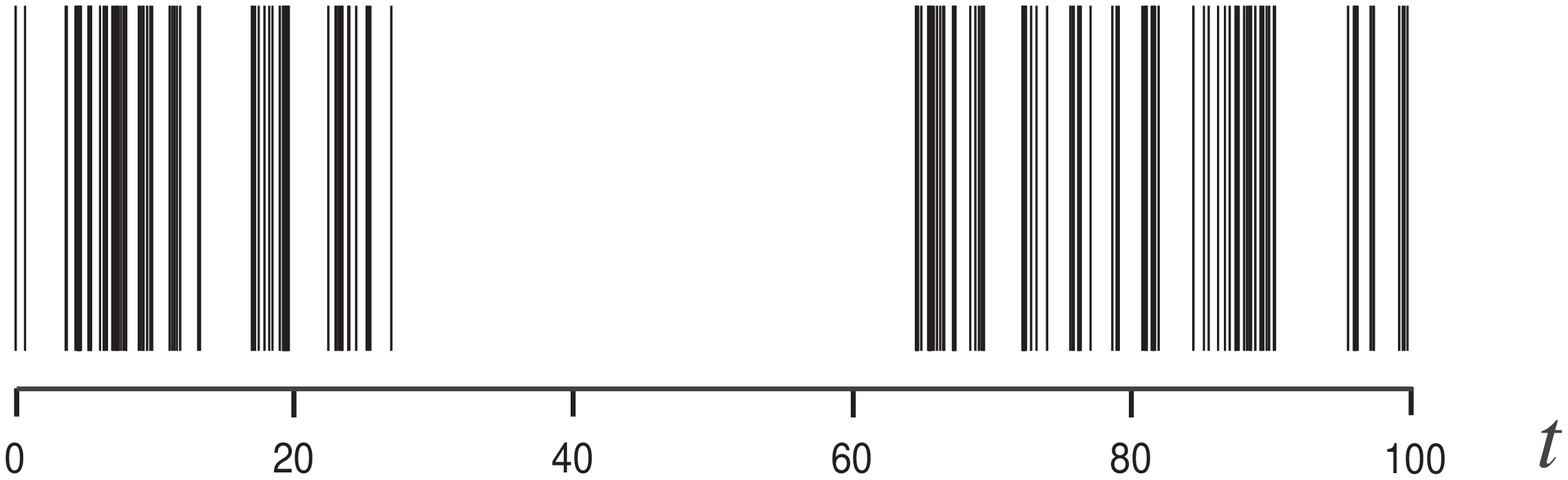}}   
   \fcaption{The operational time stemming from subordination
of the two processes depicted in Fig.1. Note that the bar-code-like set
shows both the intervals of high condensation and the long lacunae.}      
\end{center}\end{figure}

Note also that the fact that the TET and TST are not inverse of each other
is mirrored by the fact that within the formalism based on the fractional
Fokker-Planck equations (FFPE), the first one corresponds to an additional
fractional \textit{time derivative} in the l.h.s. of the FFPE, while the
second one is represented by a fractional \textit{spatial derivative, }see
Refs.\cite{Fogedby1,Hilfer,MKS,Sune}. Note also that the noncommutativity
mentioned above shows that the order of application of these derivatives is
fixed and cannot be arbitrarily changed.

\section{The ''paradoxical'' diffusion}

A process subordinated to a L\'{e}vy-CTRW under TET (a time transform
leading to subdiffusive CTRW) was considered in detail in Ref.\cite{Fogedby1}%
. We now know that this process subordinates normal random walks under a
combination of TST and TET of different indices $\beta $ and $\gamma $. The
overall behavior of the process is superdiffusive for $\gamma <\beta $ and
subdiffusive for $\gamma >\beta $. This is easy to understand since the
scaling considerations show that the operational time grows superlinearly
with physical time in the first case and that the behavior is sublinear in
the second case. Note that the index $\mu $ of the corresponding
L\'{e}vy-flight is exactly $2\beta $, so that this behavior is exactly the
one obtained in Ref.\cite{Fogedby1}. In the case $\beta =\gamma $ the
operational time grows linearly with the physical one: Ref.\cite{Fogedby1}
suggests that it falls into the diffusion universality class. On the other
hand this diffusion is a very special one: We will call the process
subordinated to RW under such operational time a paradoxical diffusion. The
random process defining an operational time stemming from a combination of
TST and TET of the same index $\gamma $ has interesting properties: $n$
typically grows proportional to $t$; on the other hand, neither a
well-defined density, nor a well-defined mean waiting-time exists.

\begin{figure} \begin{center}   
  \fbox{\epsfysize=6cm\epsfbox{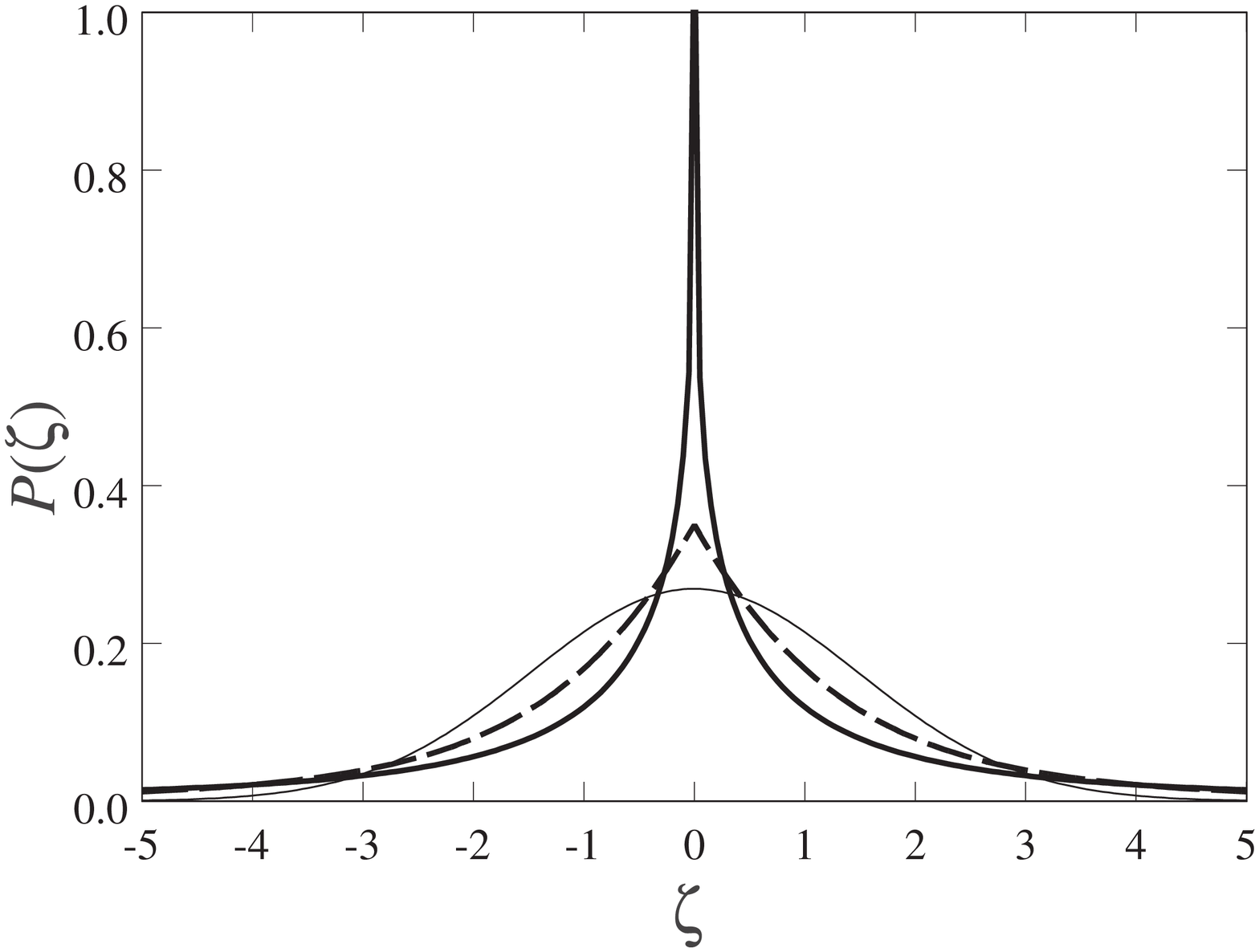}}   
   \fcaption{The PDF of the random walker's positions for
paradoxical diffusion. The PDFs are plotted as a function of dimensionless
variable $\zeta =x/Q$, where $Q$ is the position of the upper quartile of
the corresponding distribution. The thick full line corresponds to RW under
the subordination of a TST and TET, Eq.(\ref{Par05}); the dashed line
corresponds to the inverse situation, Eq.(\ref{05Par}). A thin full line
represents a Gaussian distribution of the same width.}   
\end{center}\end{figure}

Let us first discuss the situation mentioned in the beginning of the
section: a RW subordinated to L\'{e}vy-distributed operational time, driven
by a sublinear one. The PDF of the corresponding random walks has power-law
tails, namely, exactly those of a L\'{e}vy-distribution of index $\gamma $.
On the other hand, the overall width of the corresponding curve grows as $%
\Lambda \simeq \sqrt{t}$. Moreover, the whole distribution scales a as a
function of dimensionless displacement $\xi =x/\Lambda $: the overall
behavior is somewhat similar to one found on a polymer chain with bridges.
The overall form of the function can be found using the well-known
expression for $p(n,u)$, the Laplace-transform of the probability $p(n,t)$
to make exactly $n$ steps up to time $t$. Such a process corresponds to a
directed motion under the same operational time as CTRW. For the ordinary
renewal process one has $p(n,u)=\frac{1}{u}\left[ 1-\psi (u)\right] \psi
^{n}(u)$, with $\psi (u)\simeq 1-u^{\gamma }$ \cite{Blumofen}. For $u$ small
($t$ large) this form corresponds to 
\begin{equation}
p(n,u)\simeq u^{\gamma -1}\exp (-nu^{\gamma }).  \label{Pnu}
\end{equation}
Considering paradoxical diffusion as a process subordinated to
L\'{e}vy-flights of index $2\gamma $ under operational time given by $p(n,t)$%
, we get for $P(k,u)$, the Fourier-Laplace transform of $P(x,t),$ 
\begin{equation}
P_{\gamma }(k,u)=\int_{0}^{\infty }e^{-\left| k\right| ^{2\alpha
}n}p(n,u)dn\simeq \frac{u^{\gamma -1}}{\left| k\right| ^{2\gamma }+u^{\gamma
}}.  \label{Paradif}
\end{equation}
The scaling nature of the distribution is immediately evident, the nature of
its power-law tails follows from the asymptotic analysis for $k$ small: The
tail of $P_{\gamma }(\xi )$ stems from those of $L(x,2\gamma ,0)$ and has a
power-law asymptotics $P_{\gamma }(\xi )\propto \xi ^{-1-2\gamma }$ ($\gamma
<1$). Note that such a distribution was obtained in Ref.\cite{Fogedby1} as a
solution of a fractional diffusion equation, describing a random process
incorporating L\'{e}vy-jumps taking place under sublinear operational time.
As an example let us consider the distribution $P_{1/2}(x,t)$, i.e. one for $%
\gamma =1/2$. This distribution has a simple analytical form, which can be
obtained by an inverse Laplace-Fourier transformation of Eq.(\ref{Paradif}).
The inverse Laplace transform of Eq.(\ref{Paradif}) is one given in 3.21 of
Ref.\cite{Laplace}and reads: $P_{1/2}(k,t)=\exp (k^{2}t)\mbox{erfc}(\left|
k\right| t^{1/2})$. The inverse (cosine)-Fourier transform of this function
is given by No. 10.6 of Ref.\cite{Fourier} and reads: 
\begin{equation}
P_{1/2}(x,t)=-\frac{1}{2\sqrt{t}}\pi ^{-3/2}\exp (x^{2}/4t)\mbox{Ei}%
(-x^{2}/4t),  \label{Par05}
\end{equation}
where $\mbox{Ei}(x)$ is the exponential integral, see Eq. 5.1.2 of Ref.\cite
{Abrasteg}. The corresponding function is a scaling function of $\xi =$ $%
x/t^{1/2}$; its behavior for $\xi $ large follows from asymptotic expansion
of $-\mbox{Ei}(-x)={\mathrm{E}_{1}}(x)=x^{-1}e^{x}[1-1/x+...]$, so that
asymptotically $P_{1/2}(\xi )$ shows the $\xi ^{-2}$-like tail, similar to
one of Cauchy-distribution. For $\xi \rightarrow \infty $ the distribution $%
P_{1/2}(k,t)$ shows week (logarithmic) singularity (following from
Eq.(5.1.11) of Ref.\cite{Abrasteg}), a sign of strong lacunarity of the
corresponding operational time. The asymptotic analysis of Eq.(\ref{Paradif}%
) shows that such integrable singularities appear in the center of
distribution for $0<\gamma \leq 1/2$: the behavior for $\xi \rightarrow 0$
is given by $P_{\gamma }(\xi )\propto \xi ^{2\gamma -1}$, for $\gamma =1/2$ $%
P_{\gamma }(\xi )$ diverges logarithmically, as we have already seen in Eq.(%
\ref{Par05}).

The distribution $P_{1/2}(\xi )$ is plotted in Fig.3 together with the
Gaussian distribution (i.e. the distribution $P_{1}(\xi )$ of the same
class, the one corresponding to a normal diffusion) and with the
distribution stemming from the inverse order of application of TET and TST
to a simple diffusion, which is discussed in detail in the next section. All
distributions are normalized in such a way that their quartiles coincide.
Note that the quartiles of $P_{1/2}(\xi )$ are situated at $\pm 0.841$.

\section{Non-commutativity of time-subordination}

Applying the transformations other way around, i.e. considering a process
subordinated to Scher-Montroll CTRW under L\'{e}vy-time, we get a process
which is different from one discussed above. Let us start from a simple
example.

Let us note that the TET of index 1/2 (corresponding to an inverse
Laplace-transform of the function $e^{-n\sqrt{u}}/\sqrt{u}$) is given by 
\begin{equation}
Q_{1/2}(n,t)=\frac{1}{\sqrt{\pi t}}e^{-n^{2}/4t},  \label{TET1/2}
\end{equation}
i.e. corresponds to a part of a Gaussian distribution for $n>0$, so that $n$
typically grows as $t^{1/2}$. The corresponding TST is given by a
distribution, Eq.(\ref{Smir}), $R_{1/2}(T,n)=\frac{n}{2\sqrt{\pi }T^{3/2}}%
e^{-n^{2}/4T}$. The subordination of these two processes is described by a
function 
\begin{eqnarray}
S_{1/2}(T,t) &=&\int_{0}^{\infty }\frac{1}{\sqrt{\pi t}}e^{-n^{2}/4t}\cdot 
\frac{n}{2\sqrt{\pi }T^{3/2}}e^{-n^{2}/4T}dn  \nonumber \\
&=&\frac{2}{\pi t}\sqrt{\frac{t}{T}}\left( \frac{T}{t}+1\right) ^{-1},
\label{Paradox}
\end{eqnarray}
which is a probability distribution with the tail decaying as $T^{-3/2}$ (as
a tail of a stable distribution of index 1/2) and with the square-root
singularity at zero. Note that this distribution is just a solution of a
fractional Liouville equation describing directed motion under such an
operational time, just like Eq.(\ref{Par05}) is the solution of a fractional
diffusion equation. This is a process subordinated to a L\'{e}vy one under
sublinear time growth.

We now show that the $Q$- and $R$- distributions leading to the paradoxical
diffusion are not commutative: An operatational time resulting from a $R*Q$
transformation has a different distribution from one stemming from a $Q*R$%
-one. For example, the distribution $S_{1/2}(T,t)$ given by Eq.(\ref{Paradox}%
) is $S_{1/2}(T,t)=Q*R=\int Q(n,t)R(T,n)dn$. Let us calculate a conjugated
disrtibution, $S_{1/2}^{*}(T,t)=R*Q=\int R(n,t)Q(T,n)dn$, one describing a
process subordinated to a sublinear growth under the operational time
growing according to a L\'{e}vy distribution. The distribution $%
S_{1/2}^{*}(T,t)$ is given by:

\begin{eqnarray}
S_{1/2}^{*}(T,t) &=&\int_{0}^{\infty }\frac{1}{\sqrt{\pi n}}%
e^{-T^{2}/4n}\cdot \frac{t}{2\sqrt{\pi }n^{3/2}}e^{-t^{2}/4n}dn  \nonumber \\
&=&\frac{2t}{\pi }\frac{1}{t^{2}+T^{2}},  \label{Conjug}
\end{eqnarray}
i.e. corresponds to a positive part of a Cauchy-distribution. Note that even
such a robust scaling property of a probability distribution as a nature of
its power-law tail is different from one of the conjugated counterpart.

The plausible scaling consideration here is as follows. The distribution $%
Q(T,n)$ has all moments, so that for $n$ large the value of $T$ is
well-defined and is of the order of $n^{\alpha }$, $\alpha <1$. On the other
hand, the distribution of $n$ as a function of $t$ is broad and shows a
power-law tail $P(n,t)\propto t^{-1/\alpha }(n/t^{1/\alpha })^{-1-\alpha
}\propto tn^{-1-\alpha }$. Changing now variable from $n$ to $T\propto
n^{\alpha }$ we get the asymptotics of the PDF of $T\,$in a form: $%
P(T,t)\propto tT^{-2}$, independently on $\alpha $. We note thus that the
probability distribution subordinating a subliner continuous-time directed
motion under the L\'{e}vy-distributed operational time of the same index has
a power-law tail decaying as $T^{-2}$, i.e. is similar to a
Cauchy-distribution.

The process subordinated to a Gaussian RW under operational time defined by $%
S_{1/2}^{*}(T,t)$ is also not a normal diffusion, but represents a marginal
situation of a distribution whose second moment diverges logarithmically.
The corresponding PDF shows power-law tails of a $x^{-3}$ type. This PDF is
given by: 
\begin{equation}
P_{1/2}^{*}(x,t)=\int_{0}^{\infty }\frac{1}{\sqrt{2\pi T}}e^{-x^{2}/2T}\frac{2t}{\pi }%
\frac{1}{t^{2}+T^{2}}dT.
\end{equation}
Changing to a new variable $\zeta =x^{2}/2T$ and then introducing a scaling
variable $\xi =x/\sqrt{t}$ we get the PDF $P(x,t)$ as a scaling function of $%
\xi $: 
\begin{equation}
P_{1/2}^{*}(\xi )=\frac{1}{\pi ^{3/2}}\left| \xi \right| \int_{0}^{\infty }\frac{\zeta
^{1/2}e^{-\zeta }}{\zeta ^{2}+\xi ^{4}/4}d\zeta .  \label{INTot}
\end{equation}
For $\xi $ large the corresponding integral decays as $(2/\pi )\xi ^{-3}$.
Note that Eq.(\ref{INTot}) can be expressed in terms of Fresnel sine- and
cosine integrals, $S(x)$ and $C(x)$, so that $P(\xi )$ can be obtained in a
closed form: 
\begin{eqnarray}
P_{1/2}^{*}(\xi ) &=&\frac{1}{\sqrt{\pi }}\left\{ \sin \left( \frac{\xi ^{2}%
}{2}\right) \left[ 1-2S\left( \frac{\left| \xi \right| }{\sqrt{\pi }}\right)
\right] \right.   \nonumber \\
&&\left. +\cos \left( \frac{\xi ^{2}}{2}\right) \left[ 1-2C\left( \frac{%
\left| \xi \right| }{\sqrt{\pi }}\right) \right] \right\} ,  \label{05Par}
\end{eqnarray}
see Eq.(2.3.7.10) of Ref.\cite{BP}. The corresponding distribution is also
plotted in Fig.3\ \ as a dashed line. Note that the distribution shows a
cusp-singularity at $\xi =0$. The value of $P(\xi )$ in this point is $1/%
\sqrt{\pi }=0.564.$.. The quartiles of this distribution are situated at $%
\pm 0.621$.

\section{Relaxation phenomena under temporal subordination}

The fact that the L\'{e}vy dynamics can follow from a temporal subordination
is important if one wants to analyze the possible thermodynamical
implications of the L\'{e}vy-flight transport. Imagine an ensemble of
thermodynamical systems (say, Brownian particles in a harmonic potential)
which was put out of equilibrium and then let relax. As discussed in Sec.2,
such relaxation will lead to a stationary state corresponding to a normal
equilibrium Boltzmann distribution. Since this distribution is
time-independent, it would not change under temporal subordination, so that
the systems with L\'{e}vy dynamics may have very ordinary thermodynamical
equilibrium states and thus be described by normal Gibbs-Boltzmann entropy.
The non-Boltzmann nature of the equilibrium found in Ref.\cite{Sune} was
connected with the fact that the linear response was considered, as proposed
by Ref.\cite{FogQ}, an assumption at variance with the findings of Sec. 4.
Let us now discuss the relaxation to this equilibrium.

A system slightly outside of the equilibrium can be considered as evolving
under the influence of the linear restoring force. In the operational time
of the system (marked by the number $n$ of jumps) this relaxation will be
described by a Fokker-Planck equation. For an overdamped particle in a
harmonic potential we get, for example: 
\begin{equation}
\frac{\partial P}{\partial n}=\frac{\partial }{\partial x}\left( \gamma kxP+D%
\frac{\partial }{\partial x}P\right)  \label{FoPla1}
\end{equation}
Note that the values of $\gamma $ and $D$ fulfill the Einstein's relation, $%
\gamma =D/kT$. The Green's function of Eq.(\ref{FoPla1}) has a form of a
Gaussian distribution and reads:

\begin{eqnarray}
G(x,n\left| x_{0},n_{0}\right. ) &=&\sqrt{\frac{\gamma }{2\pi
D(1-e^{-2\gamma (n-n_{0})})}}   \nonumber \\
&&\times \exp \left( -\frac{\gamma k(x-e^{-\gamma (n-n_{0})}x_{0})^{2}}{%
2D(1-e^{-2\gamma (n-n_{0})})}\right) ,  \label{Green1}
\end{eqnarray}
see Sec. 5.4 of Ref. \cite{Risken}. This equation gives us e.g. the PDF at
time $n$ in a system, in which the particles were all situated at $x=x_{0}$
at $t=t_{0}$. It is easy to see that the first two central moments $%
M_{1}=\left\langle x\right\rangle $ and $M_{2}=\left\langle (x-\left\langle
x\right\rangle )^{2}\right\rangle $ relax exponentially to their equilibrium
values, so that 
\begin{equation}
\left\langle x(n)\right\rangle =x_{0}\exp (-\tau ^{-1}n)  \label{Aver}
\end{equation}
and 
\begin{equation}
\sigma ^{2}(n)=\frac{D}{k\gamma }(1-\exp (-2\tau ^{-1}n)),  \label{Disper}
\end{equation}
being a typical pattern of relaxation of a system with only one
relaxation time $\tau =(k\gamma )^{-1}$. Since all higher moments of a
Gaussian distribution are the combinations of the lower two, they also relax
to their equilibrium values in a (multi-)exponential fashion. Let us start
from the Fourier-transform of Eq.(\ref{Green1}) and to note that under
subordination

\begin{eqnarray}
P(k,t) &=&\int \exp \left( -ikx^{^{\prime }}e^{-\gamma
n}-Dk^{2}(1-e^{-2\gamma n})/2\gamma \right)   \nonumber \\
&&\times t^{-1/\alpha }L(n/t^{1/\alpha },\alpha ,-\alpha )dn.  \label{Momex}
\end{eqnarray}

Let us moreover expand the exponential term in a Taylor series in $k$: the
coefficients of this series give the moments of the corresponding
distribution. From Eq.(\ref{Momex}) it follows then that the $i$-th moment
is a combination of integrals of the type 
\begin{equation}
\Phi (t)=\int_{0}^{\infty }\exp (-\lambda n)t^{-1/\alpha }L(n/t^{1/\alpha
},\alpha ,-\alpha )dn
\end{equation}
with $\lambda =m\gamma $, $0\leq m\leq i$. Using the fact that a
Laplace-transform of a fully asymmetric L\'{e}vy-distribution is a stretched
exponential function, we get: 
\begin{equation}
\Phi (t)=\exp (-A(\lambda t^{1/\alpha })^{\alpha })=\exp (-A\lambda ^{\alpha
}t).
\end{equation}
This means that the exponential relaxation under L\'{e}vy dynamics stays a
simple exponential relaxation (only the corresponding relaxation time
changes). For example, the first moment of the distribution (the particle's
position) still relaxes exponentially to its equilibrium value of zero. On
the other hand, the dependence of the relaxation time on the outer
parameters (say, temperature) entering through the values of $\gamma $ and $%
D $ can change considerably. Thus, the superdiffusive L\'{e}vy-flights
dynamics in the force-free case can coexist with standard thermodynamics and
with very simple relaxation patterns as soon as the case of a harmonic force
is concerned.

Let us consider the relaxation in a harmonic potential under ''paradoxical''
diffusion. Here again we can use the moment expansion, Eq.(\ref{Momex}), and
put down the expression for the characteristic function of the overall
distribution:

\begin{eqnarray}
P(k,t) &=&\int \exp \left( -ikx^{^{\prime }}e^{-\gamma
n}-Dk^{2}(1-e^{-2\gamma n})/2\gamma \right)    \nonumber \\
&&\times S_{\alpha }(n,t)dn.
\end{eqnarray}
Note that the moments of the corresponding distribution are the combinations
of the functions: 
\begin{equation}
\Phi (t)=\int_{0}^{\infty }\exp (-\lambda T)S_{\alpha }(T,t)dT.
\end{equation}
Note that $S_{\alpha }(n,t)$ is a PDF of a process subordinated to a
L\'{e}vy distribution under TET: 
\begin{equation}
S_{\alpha }(T,t)=\int d\tau \tau ^{-1/\alpha }L_{\alpha }(T/\tau ^{1/\alpha
},\alpha ,-\alpha )Q_{\alpha }(\tau ,t)d\tau
\end{equation}
Thus, a Laplace transform of $S$ according to its outer time-variable is a
stretched-exponential, so that 
\begin{equation}
\Phi (t)=\int_{0}^{\infty }p(\tau ,t)\exp (-A\lambda ^{\alpha }\tau )d\tau .
\end{equation}
Let us take a Laplace-transform of this expression. Using Eq.(\ref{Pnu}) we
get: 
\begin{equation}
\Phi (u)=\int_{0}^{\infty }u^{\alpha -1}\exp (-\tau u^{\alpha })\exp
(-A\lambda ^{\alpha }\tau )d\tau =\frac{u^{\alpha -1}}{u^{\alpha }+A\lambda
^{\alpha }}.
\end{equation}

For small $u$ (long times) this corresponds to a power-law decay of $\Phi
(t) $ of a form $\Phi (t)\propto t^{-\alpha }$ for $t>>\lambda ^{-1}$. Thus,
the relaxation in the case of paradoxical diffusion resembles those in
normal CTRW and is dominated by large lacunae. In the case when the
processes are subordinated other way around, i.e. according to $S_{\alpha
}^{*}(T,t)$, the decay at longer times follows the universal $t^{-1}$-law:
for example for $\alpha =1/2$ we get: 
\begin{eqnarray}
\Phi (t) &=&\frac{2t}{\pi }\int_{0}^{\infty }\exp (-\lambda T)\frac{1}{%
t^{2}+T^{2}}dT=  \nonumber \\
&=&\frac{2\lambda}{\pi }\left[ \sin (\lambda t) {\mbox{ci}} (\lambda t)-\cos
(\lambda t) {\mbox{si}} (\lambda t)\right] ,
\end{eqnarray}
see Eq.(2.3.7.11) of Ref.\cite{BP} (here the integral sine- and
cosine-functions, ${\mbox{si}}(x)=-\int_{x}^{\infty }\frac{\sin x}{x}dx$ and 
${\mbox{ci}}(x)=-\int_{x}^{\infty }\frac{\cos x}{x}dx,$ are used). For $%
\lambda t\gg 1$ we get: 
\begin{equation}
\Phi (t)\simeq \frac{2}{\pi }(\lambda t)^{-1},
\end{equation}
which asymptotic behavior is universal for all L\'{e}vy-driven CTRWs of the
same index.

\section{Conclusions}

A broad range of physical processes can be described as processes
subordinated to a random walk under some operational time. In particular,
such subordination leads to anomalous transport properties, the well-known
example being the Scher-Montroll continuous-time random walks, a process in
which the operational time (given by the number of steps) is sublinear in
the physical time $t$. Here we have considered the processes subordinated to
a diffusive process under operational time governed by a
L\'{e}vy-distribution with index $0<\alpha <1$, namely the operational time
superlinear in physical one. We have shown that in the absence of outer
forces this subordination leads exactly to L\'{e}vy-flights. The response of
such a system to weak outer force is strongly nonlinear. Interestingly
enough the relaxation patterns in such systems are simpler than expected.
Thus, we show that the behavior in the presence of a weak harmonic force
corresponds to a simple exponential relaxation to a normal Boltzmann
distribution. The combination of super- and sublinear operational times
(i.e. L\'{e}vy-flights under sublinear operational time or the
Scher-Montroll CTRW under L\'{e}vy-time) correspond to the ''paradoxical''
diffusion, a random process which in a force-free case leads to the
probability-distributions of the particle's displacements, which show the
power-law tails and lack the second moment. The width of the distribution,
on the other hand, grows proportionally to the square-root of time, showing
a typically diffusive behavior. Some physical implications of these findings
have been discussed.

\section{Acknowledgments}

The author is indebted to S. Jespersen, Prof. A. Blumen and Prof. J. Klafter
for fruitful discussions. Financial support by the Deutsche
Forschungsgemeinschaft through the SFB 428 and by the Fonds der Chemischen
Industrie is gratefully acknowledged.

\end{document}